\documentclass{aa}  
\usepackage{graphicx}
\usepackage{txfonts}
\usepackage{lscape}
\usepackage{multirow}
\usepackage{natbib}

\begin{document}

\title{Astrometric and photometric accuracies in high contrast imaging: The SPHERE speckle calibration tool (SpeCal)\thanks{Based on observations collected at the European Organisation for Astronomical Research in the Southern Hemisphere under ESO programme 097.C-0865.}}

\author{R. Galicher\inst{1}, A. Boccaletti\inst{1}, D. Mesa\inst{2,3},
  P. Delorme\inst{4}, R. Gratton\inst{2}, M. Langlois\inst{5,6},
  A.-M. Lagrange\inst{4}, A.-L. Maire\inst{7}, H. Le Coroller\inst{5},
  G. Chauvin\inst{4}, B. Biller\inst{8},
  F. Cantalloube\inst{7}, M. Janson\inst{9}, E. Lagadec\inst{10},
  N. Meunier\inst{4}, A. Vigan\inst{5}, J. Hagelberg\inst{4},
  M. Bonnefoy\inst{4}, A. Zurlo\inst{11,12,13}, S. Rocha\inst{4},
  D. Maurel\inst{4}, M. Jaquet\inst{5}, T. Buey\inst{1}, L. Weber\inst{14}}
\institute{Lesia, Observatoire de Paris, PSL Research University, CNRS, Sorbonne Universités, Univ. Paris Diderot, UPMC Univ. Paris 06, Sorbonne Paris Cité, 5 place Jules Janssen, 92190 Meudon, France\\
\email{raphael.galicher at obspm.fr}
\and {INAF - Osservatorio Astronomico di Padova, Vicolo della Osservatorio 5, 35122, Padova, Italy}
\and{INCT, Universidad De Atacama, calle Copayapu 485, Copiap\'{o},
Atacama, Chile}
\and{Univ. Grenoble Alpes, CNRS, IPAG, F-38000 Grenoble, France}
\and{ Aix Marseille Universit\'e, CNRS, LAM (Laboratoire d'Astrophysique de Marseille) UMR 7326, 13388 Marseille, France}
\and {CRAL, UMR 5574, CNRS, Universit\'e de Lyon, Ecole Normale Sup\'erieure de Lyon, 46 All\'ee d'Italie, 69364 Lyon Cedex 07, France}
\and{Max-Planck-Institut für Astronomie, Königstuhl 17, 69117
  Heidelberg, Germany.}
\and{Institute for Astronomy, University of Edinburgh, Blackford Hill,
  Edinburgh EH9 3HJ, UK}
\and{Department of Astronomy, Stockholm University, SE-106 91
  Stockholm, Sweden}
\and{Universit\'e C\^ote d'Azur, OCA, CNRS, Lagrange, France}
\and{Universidad de Chile, Camino el Observatorio 1515, Santiago,
  Chile}
\and{N\'ucleo de Astronom\'ia, Facultad de Ingenier\'ia, Universidad
  Diego Portales, Av. Ejercito 441, Santiago, Chile}
\and{Millenium Nucleus ``Protoplanetary Disks in ALMA Early Science'',
  Chile}
\and{Geneva Observatory, University of Geneva, Chemin des Mailettes 51, 1290 Versoix, Switzerland}}

\date\today%{Received --; accepted --}

\titlerunning{Astrometry and photometry with SpeCal}
\authorrunning{Galicher et al.}

\abstract
{The consortium of the Spectro-Polarimetric High-contrast Exoplanet REsearch installed at the Very Large Telescope (SPHERE/VLT) has been operating its guaranteed
  observation time (260 nights over five years) since February 2015. The main part
  of this time (200 nights) is dedicated to the detection
  and characterization of young and giant exoplanets on wide orbits.}{The
  large amount of data must be uniformly processed so that accurate
  and homogeneous measurements of photometry and astrometry can be
  obtained for any
  source in the field.} {To complement the European Southern Observatory
  pipeline, the SPHERE
  consortium developed a dedicated piece of software to process the
  data. First, the software corrects for instrumental
  artifacts. Then, it uses the speckle calibration tool (SpeCal) to
  minimize the stellar light halo that prevents us from detecting faint
  sources like exoplanets or circumstellar disks. SpeCal is meant
  to extract the astrometry and photometry of detected
  point-like sources (exoplanets, brown dwarfs, or background
  sources). SpeCal was intensively tested to ensure the consistency
  of all reduced images (cADI, Loci, TLoci, PCA, and others) for any SPHERE
  observing strategy (ADI, SDI, ASDI as well as the accuracy of the
  astrometry and photometry of detected point-like sources.}{SpeCal is
  robust, user friendly, and efficient at detecting and
  characterizing point-like sources in high contrast images. It is
  used to process all SPHERE data systematically, and its outputs have been
  used for most of the SPHERE consortium papers to date. SpeCal is also a
  useful framework to compare different algorithms using various sets of
  data (different observing modes and conditions). Finally, our tests
  show that the extracted astrometry and photometry are accurate
  and not biased.}{}

        
    \maketitle
    
    \section{Introduction}

The Spectro-Polarimetric High-contrast Exoplanet REsearch (SPHERE) \citep{beuzit08} is a facility-class instrument at the Very Large Telescope (VLT)
dedicated to directly imaging and spectroscopically characterizing
exoplanets and circumstellar disks. It combines a high-order adaptive-optics system with diverse coronagraphs. Three instruments are available:
an infrared dual-band imager \citep[IRDIS,][]{dohlen08}, an infrared
integral field spectrometer \citep[IFS,][]{claudi08}, and a visible
imaging polarimeter \citep[Zimpol,][]{thalmann08}.

Since first light in May 2014, SPHERE has been performing well in
all observational modes, enabling numerous follow-up studies of known
sub-stellar companions to stars and circumstellar disks as well as new
discoveries \citep[e.g.,][]{boccaletti15, deboer16, ginsky16,
  lagrange16,   maireAL16a, mesa16,perrot16, vigan16, zurlo16,
  chauvin17, bonnefoy17,   feldt17, maireAL17,  mesa17,
  samland17}. The good performance to date results from the stability of the
instrument over time and a dedicated and sophisticated software for
data reduction. 

The SPHERE reduction software uses the data reduction handling
pipeline \citep[DRH,][]{pavlov08} that was delivered to ESO with the
instrument as well as upgraded tools optimized using the first SPHERE
data to derive accurate spectrophotometric and astrometric calibrations
\citep{mesa15,maireAL16b}. The software that is implemented at the
SPHERE data center \citep{delormeP17b} first assembles tens to
thousands of images or spectra into calibrated datacubes,
removing or correcting for instrumental artifacts. This includes image processing
steps such as flat-fielding, bad pixels, background, frame selection, anamorphism
correction, true north alignment, frame centring, and spectral transmission. The
outputs of these first steps are temporal and spectral sequences of images organized
in datacubes with four dimensions hereafter: coronagraphic images and the associated
point-spread functions~(PSF). Then, the software uses the speckle calibration
tool (SpeCal) written in the IDL language and described in this paper.
SpeCal was developed in the context of the SpHere INfrared survey of
Exoplanets (SHINE), which is the main part of the SPHERE guaranteed observation
time (GTO) and is now used to process all SPHERE data obtained with IRDIS
and IFS systematically. {SpeCal uses data processing algorithms proposed in
  the literature.} Data from the Zurich imaging polarimeter (ZIMPOL) will be implemented in the future.

In~section\,\ref{sec:algo}, we describe the algorithms that SpeCal uses to optimize the exoplanet detection, while in~section\,\ref{sec:astropho} we explain how SpeCal algorithms estimate the astrometry and photometry of point-like sources like exoplanets or brown dwarfs. To address the accuracies of the data reduction in terms of astrometry and photometry we use SpeCal to reduce two sequences recorded with IRDIS (section\,\ref{sec:exampleIRDIS}) and IFS (section\,\ref{sec:exampleIfs}) during the SPHERE guaranteed time observations.

\section{Calibration of the speckle pattern}
\label{sec:algo}

\subsection{Differential imaging strategies}
\label{subsec:imdiff}
Current high contrast imaging instruments dedicated to exoplanet
detection combine an adaptive optics system to compensate for the
atmospheric turbulence and coronagraphs to attenuate the flux of the
central bright source (i.e. the star). Because the adaptive optics
system is not perfect and because of aberrations in the optics
of the telescope and instrument, part of the stellar light
reaches the science detector, creating spatial interference patterns called
speckles. The speckles mimic images of off-axis point-like sources, especially
in narrow band filters. Other factors let stellar light go
  through the coronagraph preventing the detection of faint sources in the
  raw data: chromatism of the coronagraph, atmospheric dispersion,
  diffraction effects from the secondary mirror and from spiders, low
  wind effect, and so on. Hereafter, for convenience, the stellar
  speckle pattern will refer to any type of stellar light that
  reaches the detector even if it is not in the form of speckles.
 
Strategies that are routinely used to discriminate exoplanet images from
the stellar speckle pattern include angular
differential imaging \citep[ADI,][]{marois06}, dual-band imaging, and
spectral differential imaging
\citep[SDI,][]{rosenthal96,Racine99,Marois04,thatte07}, reference
differential imaging \citep[RDI,][]{beuzit97}, and polarization
differential imaging \citep[PDI,][]{baba03,baba05}. Each strategy
relies on specific assumptions about the speckle pattern, and the
efficiency of extracting the planet signal from the speckle pattern is
directly related to the strengths and limitations of these
assumptions. When using ADI, we assume most of the optical aberrations
come from planes that are optically conjugated to the pupil plane and
remain static in the course of the observation. Keeping the pupil
orientation fixed (pupil tracking mode), we record a sequence of
images that show a stable speckle pattern while the field of view,
including an off-axis exoplanet image, rotates around the central
star. Using dual-band imaging, we assume the spectrum of the star (and
so, the speckles) is different from the exoplanet spectrum. Using SDI,
we assume the speckles are induced by an achromatic optical path
difference in a pupil plane so that we can predict the evolution of
the phase aberrations that induce the speckles with wavelength. Using
RDI, we observe several similar stars with a similar instrumental
set-up assuming the speckle pattern is stable in time. Finally, the
PDI technique assumes that, unlike the star light, the light coming from
the planet is polarized. For each strategy, several algorithms exist
to process the data.

The SPHERE instrument can record coronagraphic images
simultaneously in two spectral filters using the IRDIS subsystem for
dual-band imaging \citep{vigan10} or in 39 narrow spectral channels
using the IFS for SDI \citep{zurlo14,mesa15}. During the observations,
ADI, dual-band imaging, and SDI can be used so that the SPHERE
instrument records four-dimensional datacubes, called $I(x,y,\theta,\lambda)$: two
spatial dimensions ($x$, $y$, sky coordinates), one angular dimension
($\theta$, orientation with respect to the north direction, which
evolves with time in pupil tracking mode), and one spectral dimension
($\lambda$, wavelength). In the rest of the paper, we use "spectral
channel" to refer to the spectral dimension, and "angular channel" to
refer to the angular dimension. Also, we use SDI for both SDI and
dual-band imaging.

The SpeCal tool was developed so that all the SHINE SPHERE GTO data
  can be uniformly processed. We remind readers that this tool uses data
  processing techniques that were previously proposed in the literature.
  The interest of the tool is that all these techniques have been tested
  on several datasets to ensure that all the products are consistent
  (contrast curves, measurements of astrometry, and photometry of detected
  point-like sources). SpeCal can process data recorded using ADI,
  SDI, RDI, or the combination of ADI and SDI that is called ASDI hereafter
  (see Table\,\ref{tab:acronyms_strategy}).
\begin{table}
\centering
\begin{tabular}{cl}
  ADI&Angular differential imaging\\
  SDI&Spectral differential imaging\\
PDI&Polarization differential imaging\\
RDI&Reference differential imaging\\
 ASDI&Simultaneous ADI and SDI\\
\end{tabular}
\caption{Acronyms of the strategies of observation.}
\label{tab:acronyms_strategy}
\end{table}
When SDI or ASDI is chosen, the frames
$I(x,y,\theta,\lambda)$ are spatially scaled with wavelength to
compensate for the spectral dispersion of the speckle position and
size, the reference wavelength being the shortest one. The resulting
frames are called $I_{s}(x,y,\theta,\lambda)$. The scaling changes the
spatial sampling. An inverse scaling is performed at the end of
the data processing. The ASDI option
usually minimizes the speckle pattern more efficiently but it can
strongly bias the photometry of the objects that are detected, as
demonstrated in \citet{maireAL14} and \citet{rameau15}.

Numerous algorithms were proposed to minimize the speckle pattern in
coronagraphic images so that point-like sources (e.g., exoplanets) or
extended sources (e.g., circumstellar disks) can be detected. SpeCal
offers several algorithms depending on the observing strategy:
classical ADI \citep[cADI, ][]{marois06}, classical reference
differential imaging (cRDI), subtraction of a radial profile
(radPro), locally optimized combination of images
\citep[Loci,][]{lafreniere07b}, LociRDI, template-Loci
\citep[TLoci,][]{marois14}, principal component analysis
\citep[PCA,][]{soummer12,amara12}, and classical
averaging with no subtraction (ClasImg). The objective of each algorithm,
except ClasImg, is the determination of one speckle pattern
$A(x,y,\theta,\lambda)$ that is then subtracted from
$I(x,y,\theta,\lambda)$ to obtain a datacube $R(x,y,\theta,\lambda)$
where the stellar intensity is reduced:
\begin{equation}
R(x,y,\theta,\lambda)= I(x,y,\theta,\lambda)- A(x,y,\theta,\lambda).
\label{eq:mainsub}
\end{equation}
The pattern $A$ can be a function of the spectral dimension, of the
angular dimension, and of the sky coordinates.

In the ADI cases, once the $A$ pattern is subtracted, all frames of
$R$ are rotated to align their north axis. In the SDI case, the $R$ frames
are spatially scaled to recover the initial sampling. In the ASDI
case, the $R$ frames are both spatially scaled and rotated. Then, the
frames are mean-combined to sum up the off-axis source signal. The
result is a datacube $I_{\mathrm{final}}(x,y,\lambda)$. SpeCal also
uses a median combination of the $R$ frames. Hence, there are two
final images for each spectral channel.

The following sections describe how SpeCal calculates the $A$ pattern
for each algorithm, the acronyms for which are given in Table~\ref{tab:acronyms}.
\begin{table}
\centering
\begin{tabular}{lccc}
  Acronym&Name&Strat.&Section \\
  \multirow{2}*{ClasImg}&Median/average&\multirow{2}*{All}&\multirow{2}*{\ref{subsec:avera}}\\
  &combination&&\\
  &&&\\
  cADI&Classical ADI&ADI&\ref{subsec:cadi}\\
  &&&\\
  cRDI& Classical RDI&RDI&\ref{subsec:rdi}\\
  &&&\\
  \multirow{2}*{radPro}& Subtraction of&\multirow{2}*{ADI}&\multirow{2}*{\ref{subsec:radpro}}\\
&radial profile&\\
  &&&\\
  \multirow{2}*{Loci}&Locally optimized&\multirow{2}*{All}&\multirow{2}*{\ref{subsec:loci}}\\
&combination of images&\\
  &&&\\
  Tloci&Template-Loci&All&\ref{subsec:tloci}\\
  &&&\\
  LociRDI&Loci using only RDI&RDI&\ref{subsec:locirdi}\\
  &&&\\
  \multirow{2}*{PCA}&Principal component&\multirow{2}*{All}&\multirow{2}*{\ref{subsec:pca}}\\
&analysis&
\end{tabular}
\caption{Acronyms and names of algorithms used to process the data recorded using a given strategy of observation (third column). If the third column shows 'all', it means the algorithm can be used on ADI, or SDI, or ASDI data.}
\label{tab:acronyms}
\end{table}

\subsection{Classical averaging}
\label{subsec:avera}
SpeCal can provide the average and median combinations of the
rotated $R$ frames using $A=0$. This algorithm (ClasIm) can be
useful to optimize the signal-to-noise ratio of detections in
parts of the image that are dominated by background instead of
speckles.

\subsection{cADI}
\label{subsec:cadi}
In classical ADI, SpeCal averages the cube of frames over the angular
dimension for each spectral channel ($\lambda$):
\begin{equation}
A_a(x,y,\lambda)= <(I(x,y,\theta,\lambda)>_\theta.
\label{eq:Aave}
\end{equation}
Equation\,\ref{eq:mainsub} is applied and the $R$ frames are rotated and then
averaged to produce one final image $I_{\mathrm{final, a}}$ per
spectral channel.

SpeCal can also remove the median of the frames instead of the average:
\begin{equation}
A_m(x,y,\lambda)= (\mathrm{median}(I(x,y,\theta,\lambda))_\theta.
\end{equation}
Then, it uses $A_m$ in Eq.\,\ref{eq:mainsub}, rotates the $R$ frames,
and then applies a median combination. The final image
$I_{\mathrm{final, m}}$ is less sensitive than $I_{\mathrm{final, a}}$
to uncorrected hot or bad pixels. It is however harder to accurately
retrieve the photometry of a detected off-axis source from
$I_{\mathrm{final, m}}$ than from $I_{\mathrm{final, a}}$ (section\,\ref{sec:astropho}).

\subsection{cRDI}
\label{subsec:rdi}
The reference differential imaging (RDI) is especially useful to
obtain images of extended sources like circumstellar disks or to probe
small angular separations to the star because it is not subject to
self-subtraction unlike the other algorithms. The classical RDI (cRDI)
is similar to cADI but it uses $N$ reference frames
$I_R(x,y,n,\lambda)$ to derive the speckle pattern $A$, where
$n=1..N$. These reference frames are images of stars other than
the one of interest ($I_R\ne I$). The resulting
$A$ pattern is
\begin{equation}
A(x,y,\lambda)= (\mathrm{median}(I_R(x,y,n,\lambda))_n.
\end{equation}

\subsection{Subtraction of a radial profile}
\label{subsec:radpro}
For detecting and studying extended sources, SpeCal proposes the
radPro algorithm that first works out the average of the datacube over
the angular dimension\,(Eq.\,\ref{eq:Aave}). Then, it calculates the
azimuthally averaged profile in rings of one pixel width. Finally,
the $A$ pattern is the centro-symmetrical image that is derived from
this profile. 

\subsection{Loci}
\label{subsec:loci}
\begin{figure}[!ht]
  \centering
  \includegraphics[width=5cm]{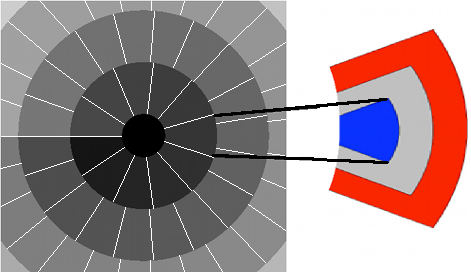}
  \caption{Loci and TLoci regions of interest (left figure and
    central blue region in the right figure) and TLoci optimizing
    region (exterior red region in the right figure).}
  \label{fig:tlociregion}
\end{figure}

The SpeCal Loci algorithm is described in \citet{lafreniere07b}. For a
given $\theta$, a given $\lambda$, and a given region in the field
(blue in Fig.\,\ref{fig:tlociregion}), the algorithm calculates the linear
combination of the other frames $I_s(x,y,\theta',\lambda')$
($\theta'\ne\theta$ and $\lambda'\ne\lambda$ in ASDI) to build
$A(x,y,\theta,\lambda)$ that minimizes $|R|$ (Eq.\,\ref{eq:mainsub})
in the region of interest.
In addition to the algorithm of \citet{lafreniere07b}, we impose the coefficients of the linear
combination to be positive. Conversely to \citet{lafreniere07b}, the
region of interest where $A$ is applied and the region of optimization
from which $A$ is derived are the same in SpeCal. They are defined
using:
\begin{itemize} 
\item the radial width of the region \citep[d$r$ in][]{lafreniere07b};
\item the number of PSFs inside the region \citep[$N_A$ in][]{lafreniere07b}.
\end{itemize}

We use different parameters to \citet{lafreniere07b} to select the
frames that are used in the linear combination.
%In SpeCal, we set:
%\begin{itemize}
%\item the number of frames that are used to build $A$: the most
%  correlated frames in each region are selected (the frames can be
%  different from one region to another).
%\item the minimum throughput $\tau$ of a putative off-axis source in
%  each region accounting for self-subtraction. This parameter can be
%  linked to $\delta_{\mathrm{min}}$ of \citet{lafreniere07b}.
%\end{itemize}
Consider the frame $I_s(x,y,\theta, \lambda)$ (ASDI case, section\,\ref{subsec:imdiff}). First, we assume the image $I_p$ of a
putative off-axis source in $I_s$ is the two-dimensional Gaussian function with the
full width half-maximum ({\sc fwhm}) estimated from the
recorded PSF (see section\,\ref{subsec:commonoutputpsf}). If the source is
in the region centered on $(x_0,y_0)$ in
$I_s(x,y,\theta, \lambda)$, it is angularly shifted by
$s_\theta=r_0\,(\theta_i-\theta)$ and radially shifted by
$s_r=r_0\,(1-\lambda/\lambda_j)$ in $I_s(x,y,\theta_i,\lambda_j)$, where
$r_0 = \sqrt{x_0^2+y_0^2}$
is the angular separation from the star in {\sc fwhm} unit. If only
$I_s(x,y,\theta_i, \lambda_j)$ was used to build $A$,
the normalized intensity of the off-axis source integrated within a
disk of one-{\sc fwhm} diameter in the frame $R$ would be
\begin{equation}
  \tau' = \frac{\iint_{r<\textrm{\sc fwhm}/2}
  \left[I_s(x',y',\theta, \lambda)-I_s(x',y',\theta_i,
    \lambda_j)\right]\,\mathrm{d}x'\,\mathrm{d}y'}
        {\iint_{r<\textrm{\sc fwhm}/2} I_p(x',y',\theta, \lambda)
          \,\mathrm{d}x'\,\mathrm{d}y'}
\end{equation}
using $x'=x-x_0$, $y'=y-y_0$ and $r=\sqrt{x'^2+y'^2}$. Then,
\begin{equation}
  \tau' = 1- \frac{\iint_{r<\textrm{\sc fwhm}/2}
  \left[I_s(x',y',\theta_i,\lambda_j)\right]\,\mathrm{d}x'\,\mathrm{d}y'}
        {\iint_{r<\textrm{\sc fwhm}/2} I_p(x',y',\theta, \lambda)
          \,\mathrm{d}x'\,\mathrm{d}y'}.
\end{equation}
Using the radial and angular shifts $s_r$ and $s_\theta$ of
$I_s(x,y,\theta_i, \lambda_j),$ we find
\begin{equation}
\tau' = 1 -
\frac{E(s_\theta)\,E(s_r)}{4\,\mathrm{erf}^2\left(\alpha/2\right)},
\label{eq:tau}
\end{equation}
where $\alpha$ equals $2\,\sqrt{\ln{2}}$, and $\mathrm{erf}$ is the
error function, and
\begin{equation}
E(s) = \mathrm{erf}\left(\alpha\,\left(s+\frac{1}{2}\right)\right) -
\mathrm{erf}\left(\alpha\,\left(s-\frac{1}{2}\right)\right).
\end{equation}
Here, we assume the angular motion is linear, which is a good
assumption because $E(s)$ quickly decreases. The function $\tau'$ goes
from $0$ ($\theta=\theta_i$ and $\lambda=\lambda_j$, total
self-subtraction) to $1$ (no self-subtraction). In SpeCal, we set a
parameter $\tau$ that can be linked to the minimum motion
$\delta_{\mathrm{min}}$ of the putative off-axis source in
\citet{lafreniere07b}. If $\tau'$ is smaller than $\tau$, the frame
$I_s(x,y,\theta_i, \lambda_j)$ is rejected. Doing so for each $\theta_i$
and $\lambda_j$, we obtain a series of frames $\{I_s\}$ that
individually leave at least $\tau$ times the initial flux of the
putative source in the $R(x,y,\theta,\lambda)$. Then, we select from
$\{I_s\}$ the $N$ most correlated frames ($N$ is adjustable) to
$I_s(x,y,\theta,\lambda)$ in the considered region. We obtain the
linear combination of the $N$ frames (i.e., the $A$ pattern) that
minimizes the residual energy in this region using the
bounded-variables least-squares algorithm by
\citet{lawson95}. Finally, as the coefficients of the linear
combination are positive, the flux of the putative source in the final
image $I_{\mathrm{final}}$ is at least $\tau$ times the initial flux
for any region.

The Loci algorithm can be used to reduce ADI, SDI, or ASDI data. In
the ADI case, the Loci $A$ pattern that is worked out for a given
frame $I(x,y,\theta,\lambda)$ uses the other frames taken in the same
spectral channel only. In the SDI case, it uses the frames taken with
the same angle $\theta$ only. Finally, in the ASDI case, all frames
are spatially scaled with wavelength, and they are all used to
determine the $A$ pattern. 

\subsection{TLoci}
\label{subsec:tloci}
The SpeCal TLoci algorithm is derived from the one described in
\citet{galicher11c} and \citet{marois14} assuming a flat planet spectrum in
contrast. The parameters that are used to select the frames ($\tau$
and $N$) and to describe the regions of interest (d$r$ and $N_A$)
where the $A$ pattern is applied (blue region in
Fig.\,\ref{fig:tlociregion}) are the same as for the Loci case
(section\,\ref{subsec:loci}). The difference from Loci is the region where
$A$ is optimized (red region). In SpeCal, the gap between this region
and the region of interest is set to $0.5$ {\sc fwhm}. Hence, the
optimizing region is close enough to the region of interest so that
$A$ is efficiently optimized, and the optimizing region is far enough
from the region of interest so that the flux of a source in the latter
does not bias the linear combination. Moreover, the internal radii of
the two regions are the same in SpeCal. Finally, an additional
parameter sets the radial width of the optimizing region.
As for Loci, TLoci can be associated to ADI, SDI, and ASDI.

\subsection{LociRDI}
\label{subsec:locirdi}
A reference differential imaging algorithm using Loci is also
implemented in SpeCal. It works as described in~section\,\ref{subsec:loci}
but the frames that are used to build~$A$ are reference frames as in
the cRDI case (section\,\ref{subsec:rdi}).

\subsection{PCA}
\label{subsec:pca}
For historical reasons, two PCA algorithms are implemented in
SpeCal. The first version can be applied on IRDIS or IFS data using
the ADI or ASDI options. This algorithm follows the equation of
\citet{soummer12}. For each frame $I(x,y,\theta,\lambda)$, we subtract
its average over the field of view:
\begin{equation}
  I_z(x,y,\theta,\lambda)= I(x,y,\theta,\lambda)-
  <I(x,y,\theta,\lambda)>_{x,y}.
  \label{eq:PCAsub}
\end{equation}
In the ADI case, the principal components are calculated for
each spectral channel independently. Each frame
$I_z(x,y,\theta,\lambda)$ is then projected onto the $N$ first
components to obtain the $A$ pattern that is used in
Eq.\,\ref{eq:mainsub}, replacing $I$ with $I_z$. The $N$ parameter is
called "number of modes" hereafter. Finally, the averages that were
removed (Eq.\,\ref{eq:PCAsub}) are added back to obtain the $R$ frames
of Eq.\,\ref{eq:mainsub}. In the ASDI case, the algorithm is the same
but it works on $I_s$ instead of $I$. We note that here there is no
frame selection to minimize the self-subtraction of point-like sources
when deriving the principal components.

The second version of PCA that is implemented in SpeCal is very
similar to the first one but it can be applied on IFS data using the
ASDI option only \citep{mesa15}. The two PCA versions were tested on a
large amount of SPHERE data and they provide a very similar performance.

\subsection{Common outputs}
\label{subsec:commonoutput}

\subsubsection{Model of unsaturated PSF}
\label{subsec:commonoutputpsf}
SpeCal produces common outputs whatever the chosen algorithm. First,
it records the final images $I_{\mathrm{final}}$ normalized to the
estimated maximum of an unsaturated non-coronagraphic stellar image
with the same exposure time. Hence, the values of the pixels in
$I_{\mathrm{final}}$ give the contrast ratio to the star maximum. For
each spectral channel, the maximum $\beta$ of the stellar
non-coronagraphic image is derived from the best fit of the input PSFs
by the function
\begin{equation}
PSF_{\mathrm{model}} = \alpha+\beta\,\exp{\left(-\left(\frac{2\,
    \sqrt{(x-x_{\mathrm{PSF}})^2
      +(y-y_{\mathrm{PSF}})^2}}{\gamma}\right)^\eta\right)},
  \end{equation}
 where $\alpha$ is the background level, $\beta$ is the star maximum,
$\gamma$ and $\eta$ are related to the spatial extension of the PSF
(and to {\sc fwhm}), and $(x_{\mathrm{PSF}}, y_{\mathrm{PSF}})$ give
the center of the PSF. All these parameters are fitted accounting for
the photon noise in the provided PSF images. The SPHERE PSFs are
usually close to two-dimensional Gaussian functions ($\eta=2$) but can deviate from
them. In the context of SHINE, PSFs are usually recorded before and
after the coronagraphic sequence. SpeCal calculates the best fit to
each of the PSFs ($\sim50x50$ pixels) and runs two tests on the time series of fitted
parameters. First, it works out the average and standard deviation of
the normalization factor $\beta$ over time, records the two values,
and sends a warning if the flux of the star $\beta$ varies by more
than $20\,\%$ between distinct PSF observations of the target. Then,
it sends a warning if the background level $\alpha$ is larger than
$10\,\%$ of $\beta$. These values ($10\,\%$ and $20\,\%$) were defined
as quality requirements based on our experience with the instrument
and the analysis of hundreds of datasets. Finally, SpeCal estimates and
records the PSF {\sc fwhm}.

\subsubsection{Calibration of photometry}
\label{subsec:commonoutputphoto}
For algorithms that bias the photometry of off-axis sources, SpeCal
estimates the throughput of the technique at each position in the
field. For cADI, radPro, and PCA, SpeCal creates a datacube of fake
planets that are on a linear spiral centered on the star (one planet per
$2 $\,{\sc fwhm}). For each planet, we use the recorded PSF and its
flux equals ten times the local residual flux in
$I_{\mathrm{final}}$. The fake planet datacube is added to the
datacube $I(x,y,\theta,\lambda)$. Then, SpeCal combines the frames as
the $I(x,y,\theta,\lambda)$ were combined to get
$I_{\mathrm{final}}$. For each planet, the ratio of the flux in the
resulting image to the flux of the fake planet is calculated to obtain
the 1D-throughput as a function of the angular separation. For Loci
and TLoci, the throughput $\tau_R$ is estimated in each frame $R$ as
the average of all $\tau'$ of Eq.\,\ref{eq:tau}, weighting
$E(s_\theta)$ and $E(s_r)$ by the coefficients used to obtain
$I_{\mathrm{final}}$. We average all $\tau_R$ to obtain the
1D-throughput as a function of the angular separation. Finally, the
throughput map $T$ is the centro-symmetrical image created from the
1D-throughput. SpeCal also calculates the throughput-corrected final
image $I_{\mathrm{final}}/T$.

\subsubsection{Signal-to-noise and detection maps}
\label{subsec:commonoutputsnr}
For each spectral channel, the image~$I_{\mathrm{final}}/T$ is divided into annulii of~$0.5$\,{\sc fwhm} width. Then, in each annulus (i.e., at each angular separation), we calculate the standard deviation that is set to be the~$1\,\sigma$ contrast.
Signal-to-noise maps are also created. Each pixel gives the ratio of
the flux in $I_{\mathrm{final}}/T$ to the standard deviation of
$I_{\mathrm{final}}/T$ calculated in annulii of $1 $\,{\sc fwhm}
centered on the star. This correction is valid only for point-like
sources. For example, this correction is not valid for an extended
source like a disk.
Finally, SpeCal also provides local detection maps giving the local
standard deviation in boxes of $2 $\,{\sc  fwhm} radial size and of a
total area of $5 $\,{\sc fwhm$^2$}.

\section{Astrometry and photometry of point-like sources}
\label{sec:astropho}
All algorithms, except RDI and ClasImg, distort the image of an off-axis
point-like source in $I_{\mathrm{final}}$
(section\,\ref{subsec:planetimage}). Thus, the estimation of the position
and flux of such a source cannot be done directly from
$I_{\mathrm{final}}$
\citep{marois10b,lagrange10,galicher11c,maireAL14,rameau15}. However,
SpeCal can fit a model of an off-axis source image to the detected
source in $I_{\mathrm{final}}$ (section\,\ref{subsec:fit1}), or it can
inject a negative point-like source into the initial datacube $I$ and
adjust the position and flux of this negative source to locally
minimize the flux in $I_{\mathrm{final}}$ (section\,\ref{subsec:fit2}).

\subsection{Planet image}
\label{subsec:planetimage}
Say there is a planet whose intensity is described by $I_p(x,y,\theta,\lambda)$ whereas the stellar intensity is $I_*(x,y,\theta,\lambda)$. The $A$ pattern is derived from the cube
\begin{equation}
  I(x,y,\theta,\lambda)=I_*(x,y,\theta,\lambda)+I_p(x,y,\theta,\lambda)
\end{equation}
and part of the $A$ pattern is composed of planet signal. For example,
when using cADI, Loci, TLoci, or radPro on ADI, SDI, or ASDI data, the $A$
pattern can be expressed as
\begin{eqnarray}
  A(x,y,\theta,\lambda) =& \sum_i\sum_j c_{i,j}\,I(x,y,\theta_i,\lambda_j),\label{eq:A0}\\
    A(x,y,\theta,\lambda) =& \sum_i\sum_j \left[c_{i,j}\,I_*(x,y,\theta_i,\lambda_j)+ c_{i,j}\,I_p(x,y,\theta_i,\lambda_j)\right],\\
        A(x,y,\theta,\lambda) =& A_*(x,y,\theta,\lambda)+A_p(x,y,\theta,\lambda),
\label{eq:A}
\end{eqnarray}
where coefficients $c_{i,j}$ are real numbers that can be a function of
$(x,y)$, and
\begin{equation}
  \left\{
\begin{array}{l}
  A_*(x,y,\theta,\lambda)=\sum_i\sum_j c_{i,j}\,I_*(x,y,\theta_i,\lambda_j)\\
  A_p(x,y,\theta,\lambda)=\sum_i\sum_j c_{i,j}\,I_p(x,y,\theta_i,\lambda_j).
\end{array}
\right.
  \label{eq:Ap}
\end{equation}
Hence, the $A$ pattern is contaminated by planet signal and, when subtracting the $A$ pattern from the initial frames $I$, part of the planet signal self-subtracts. The $R$ frames is then
\begin{equation}
R = (I_*- A_*) + (I_p-A_p),
\end{equation}
where all terms depend on $x$, $y$, $\theta$, and $\lambda$. A perfect algorithm -- that does not exist -- would be such that $A_*=I_*$ and $A_p=0$.

As the planet image moves in the field from one frame to another
(radially for SDI or azimuthally for ADI), the subtracted signal is
shifted with respect to the astrophysical position of the planet. This
results in a positive-negative pattern of the planet in each frame of
$R$ and as a consequence in the final image $I_{\mathrm{final}}$ (left
in~Fig.\,\ref{fig: template}). This pattern is not always centered on
the planet position.
The distortions of the planet image can
be minimized by carefully selecting the frames that are used to build
$A$ (so that $A_p\rightarrow0$) but it usually reduces the efficiency
of the speckle attenuation at the same time (increasing $|I_*-A_*|$).
In the case of PCA algorithms, $I(x,y,\theta_i,\lambda_i)$ in
Eq.\,\ref{eq:A0} is replaced by the principal components, which are also contaminated by the planet signal.

\subsection{Model of planet images}
\label{subsec:fit1}
\begin{figure}[!ht]
  \centering
\includegraphics[width=0.8\linewidth]{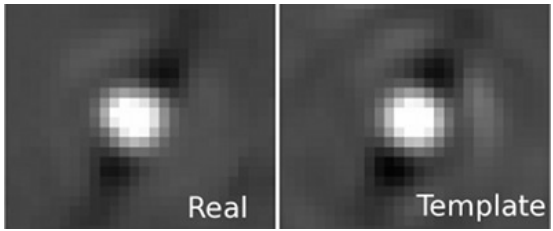}
\caption{Real~(left) and estimated~(right) images of an off-axis
  source in an ADI reduced image showing the negative wings due to
  self-subtraction of the planet flux.} 
\label{fig: template}
\end{figure}

The first way of extracting the astrometry and the photometry of an
off-axis point-like source like a planet in an ADI, SDI, or ASDI
  reduced image consists of building a model of the planet image
\citep{galicher11c}. First, we estimate the position of the detected
source in $I_{\mathrm{final}}$ with a pixel
accuracy. Then, we use the measured stellar PSF to build a sequence of
frames $I_{fp}(x,y,\theta,\lambda)$ with only one fake planet image at
the rough position accounting for the field-of-view rotation. We do
not account for the smearing that affects images. This effect may be
non-negligible at the edges of the IRDIS images if the field-of-view
rotation is fast and the exposures are long, which is rare. We combine
the frames $I_{fp}$ in the same way the frames $I$ were combined using
the same $c_{i,j}$ and we get an estimation of $A_p$
(Eq.\,\ref{eq:Ap}). Rotating (ADI case), spatially scaling (SDI case),
and averaging the frames result in a model of the planet image (right
in~Fig.\,\ref{fig: template}). The planet image has negative wings
  due to the self-subtraction of the planet flux (section\,\ref{subsec:planetimage}).
 
Then, the flux and the position of this synthetic image are adjusted
to best fit the real planet image within a disk of diameter
$3 \,${\sc  fwhm} so that it includes the positive and the negative
parts of the image. The optimization is done using
$I_{\mathrm{final}}$ , which is derived from the average of the $R$
frames after rotation and spatial scaling (and not the one obtained
using the median-combination that does not preserve
linearity). Rigorously, we should calculate the synthetic image each
time we test a new planet position. To optimize computation time, we
shift the synthetic planet image that was obtained with the rough
position. We noticed no significant difference as long as the shifts
are smaller than $\sim1 \,${\sc  fwhm}.

Once the optimization is done, we look for the excursion of each
parameter that increases the minimum residual level by a factor of
$1.15$. We set these excursions to be the~$1\,\sigma$ accuracies due to the fitting errors in the SpeCal outputs. The value
of $1.15$ is empirical but it was tested on numerous cases (high or low
signal-to-noise detection, strong or weak negative wings,
Loci, TLoci, PCA, and others). Another SpeCal output is the standard deviation
in time of the averaged flux in the coronagraphic images. To avoid
saturated parts of the images and background dominated parts, the
averaged flux is calculated inside an annulus centered on the star and
going from 30\,pixels to 50\,pixels in radius.

SpeCal can use this technique to extract astrometry and
spectro-photometry in images obtained with any algorithm
(cADI/radPro/Loci/TLoci/PCA/averaging) and any of ADI, SDI, or
ASDI. It can also be used on cRDI, LociRDI, and ClasImg final images. In
these cases, the planet image model is the stellar PSF shifted at the
position of the detection with no negative wings as the $A_p$ pattern
is null (classical fit of a non-coronagraphic image).

\subsection{Negative planets}
\label{subsec:fit2}
Another technique -- the fake negative planet --
\citep{lagrange10,chauvin12} is implemented in
SpeCal to retrieve the photometry and astrometry of point-like
sources. First, we build the sequence of frames $I_{fp}$ with the fake
planet only, as done in section\,\ref{subsec:fit1}. Then, we subtract this
datacube from the initial data $I$. We apply the algorithm and get the
final image from the $I-I_{fp}$ frames. In this final image, we
measure the standard deviation of the residual intensity inside a disk
of $3$\,{\sc fwhm} diameter centered on the rough position of
the detected planet. Modifying the fake planet position and flux, we
minimize the residual intensity. The uncertainties on the best values
are estimated as the ones in section\,\ref{subsec:fit1}. The negative
planet technique is more time-consuming than the model of planet image
technique because it calculates the $c_{i,j}$ for all the tested fake
planet positions and fluxes. It is, however, needed in some cases for
which the model of planet image technique is biased
(section\,\ref{subsec:astroIRDIS}).

\section{Reduction of IRDIS data}
\label{sec:exampleIRDIS}
In section\,\ref{sec:exampleIRDIS} and \ref{sec:exampleIfs}, we use SpeCal to
reduce two datasets as examples. In section\,\ref{sec:exampleIRDIS}, we consider
one sequence recorded during the SPHERE GTO on 2016 September 16
observing HIP2578 in IRDIS H2/H3 mode. There are 80 images of $64$\,s
exposure time and the field of view rotates by 31.5 degrees. The
seeing was about 0.5\,arcsec and the average wind speed was
7.4\,m.s$^{-1}$.

All algorithms are applied on the same datacube provided by the first
part of the SPHERE pipeline \citep[background, flat-fielding, bad
  pixels, registration, wavelength
  calibration, astrometric
  calibration;][]{pavlov08,zurlo14,maireAL16b}. The datacube is a $1024
\times 1024 \times 80 \times 2$ array. The last dimension stands for
the two spectral channels (H2 and H3). We also added three fake
planets to the data (see Table \ref{tab:fakepla}) using the recorded
stellar PSFs. We chose the position and the flux to have three typical
cases.  Planets~1 and 2 are in the speckle dominated part of the image,
planet~2 being surrounded by brighter speckles. Planet~3 is in a
region dominated by the background and not by speckles.
\begin{table}
\centering
\begin{tabular}{lccccc}
  Id&$\Delta$RA(pix)&$\Delta$DEC(pix)&Sepa(mas)&band&$C\times$1e6\\
  1&52.90&2.70&649.4&H2&3.000\\
  &&&&H3&2.400\\
  2&-25.80&8.40&318.0&H2&6.000\\
  &&&&H3&3.000\\
  3&-72.00&-71.20&1250.1&H2&1.000\\
  &&&&H3&3.000
\end{tabular}
\caption{Separations to the central star in pixels
  towards east ($\Delta$RA) and north ($\Delta$DEC), angular
  separations in mas and flux ratio with respect to the star ($C$) for
  each fake planet added to the IRDIS data.}
\label{tab:fakepla}
\end{table}

\subsection{Calibration of the speckle pattern}
\begin{figure*}
\centering
  \includegraphics[width=.9\textwidth]{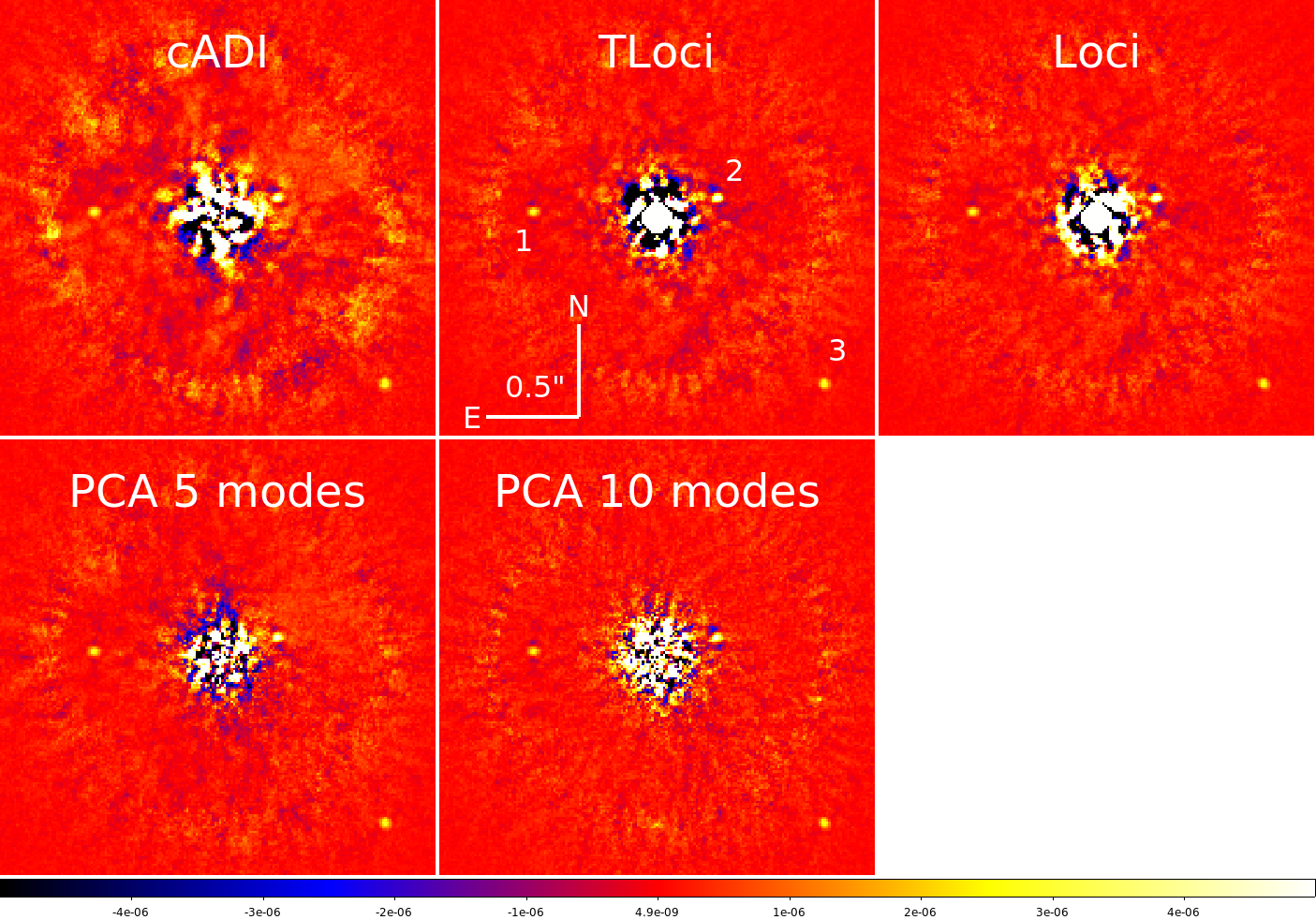}
  \caption{IRDIS example: Final images using cADI, Tloci, Loci, PCA
    (5 and 10 modes). Images are corrected from the technique
    throughput and from the coronagraph transmission. The color scale,
    which is the same for all images, shows the contrast to the star
    ratio. The spatial scale is the same for all images.}
  \label{fig:im}
\end{figure*}

\begin{figure}[!ht]
\centering
  \includegraphics[width=7.4cm]{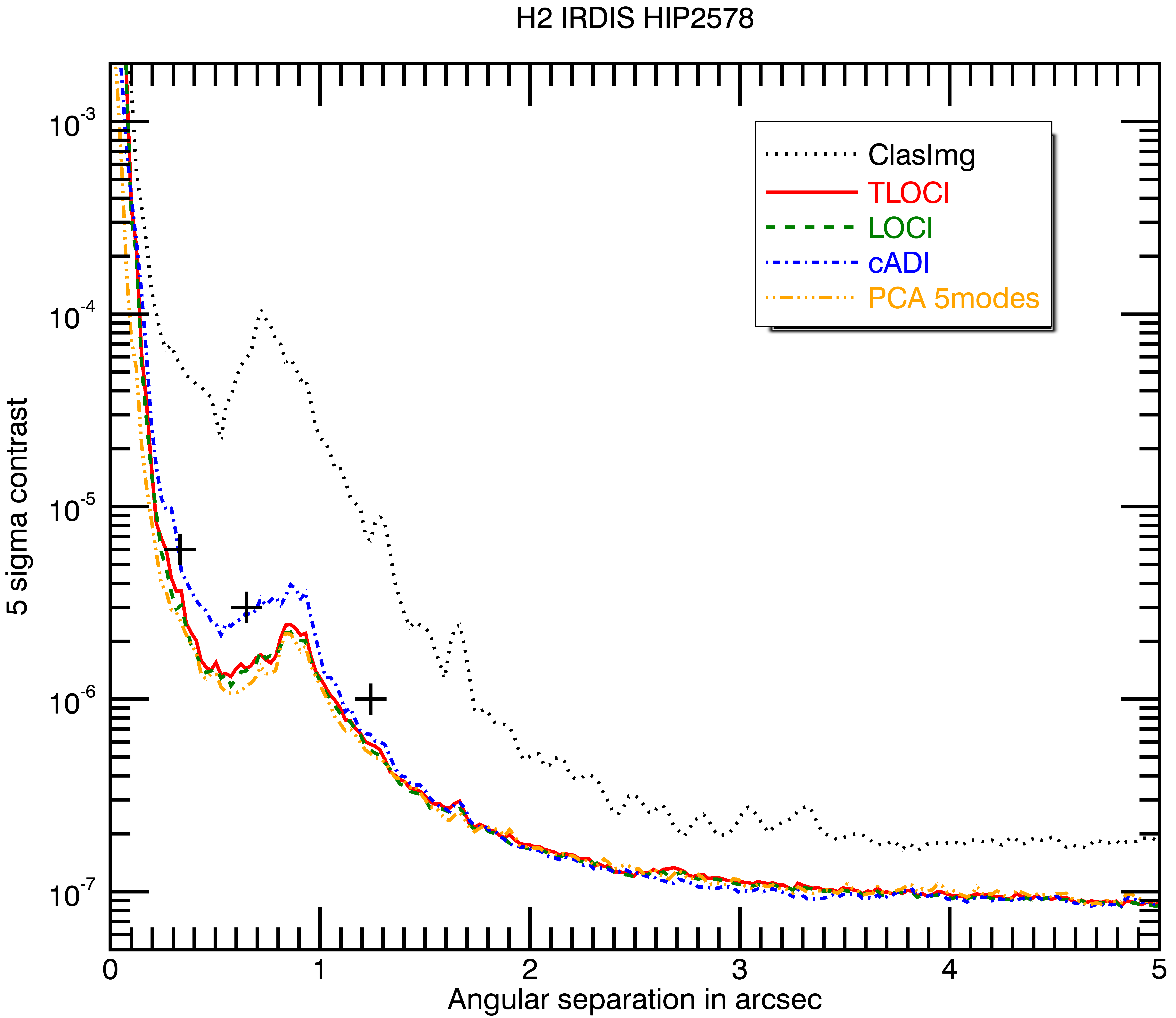}
  \caption{Contrast curves at $5\,\sigma$ before (ClasImg) and after
    minimization of the stellar light using different algorithms (cADI,
    Tloci, Loci, PCA 5 modes) on H2 data. The curves are corrected
    from the technique attenuation and from the coronagraph
    transmission. The fake planets are represented by plus symbols.}
  \label{fig:curve159}
\end{figure}

\begin{figure}[!ht]
\centering
  \includegraphics[width=7.4cm]{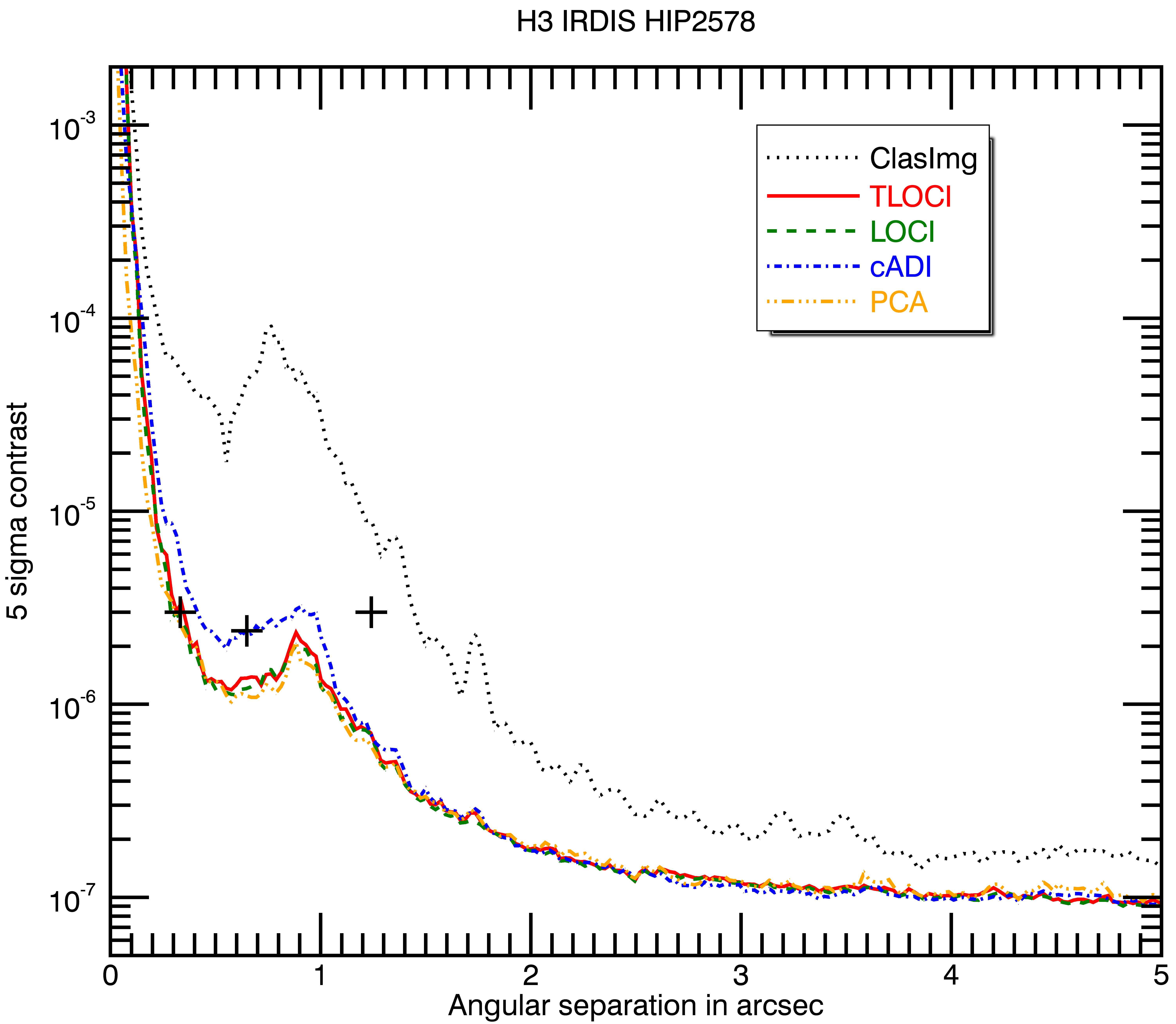}
  \caption{Same as Fig.\,\ref{fig:curve159} for H3-band.}
  \label{fig:curve167}
\end{figure}

  We apply several algorithms to the two spectral channels independently
  (no use of SDI) and we show the average of the two final images
in~Fig.\,\ref{fig:im} for cADI, TLoci, Loci, and two PCA (5 and 10
modes). Images are corrected from the technique throughput $T$
(self-subtraction of a putative planet,
section\,\ref{subsec:commonoutputphoto}) and from the coronagraph
transmission.
All images provide similar sensitivities except for cADI, which
is less efficient inside the AO correction area that is dominated by
speckles. The three fake planets are well detected in all
images. The corresponding $5 \sigma$ contrast curves are plotted
in~Figs.\,\ref{fig:curve159} (H2) and \ref{fig:curve167} (H3), where the
three planets are represented by plus symbols. The curves are
corrected for the throughput $T$ of each technique and from the
coronagraph transmission \citep{guerri11}. The latter is a function of
angular separation and it was calibrated at the telescope and via
numerical simulations.
We also overplot the contrast before any a posteriori speckle
minimization with a dotted line (ClasImg). The utility of a speckle
calibration during the data processing is obvious since none of the
planets are detected in the images with no subtraction (ClasImg).
In this example, all the
algorithms except cADI give similar performances in terms of contrast
level. For some sequences, one algorithm reaches better contrast
levels but there is no algorithm in SpeCal that is always better than
the others.

The small sample statistic bias \citep{mawet14} that mainly affects
separations at less than $0.2''$ will be implemented in the next
version of the tool. This correction will affect all the algorithm
reductions the same way.

\subsection{Measurements of astrometry and photometry}
\label{subsec:astroIRDIS}
We extract the astrometry and photometry for each detected
point-like source using the technique of the model of planet image
(section\,\ref{subsec:fit1}). The estimated contrasts to the star are
gathered in Table\,\ref{tab:astrophoto} with the associated $1\,\sigma$
uncertainties.
\begin{table*}
\centering
\begin{tabular}{ccccc|cc|cc}
  Algorithm&Id&band&$\Delta$RA(pixel)&E&$\Delta$DEC(pixel)&E&C$\times$1e6&E\\
   \hline
 \multirow{6}*{cADI}&\multirow{2}*{1}&H2&$52.70\pm0.21$&$0.95$&$2.54\pm0.20$&$0.80$&$2.775\pm0.206$&$1.09$\\
  &&H3&$53.02\pm0.22$&$0.55$&$2.48\pm0.22$&$1.00$&$2.272\pm0.163$&$0.79$\\
  &\multirow{2}*{2}&H2&$-25.94\pm0.42$&$0.33$&$8.60\pm0.37$&$0.54$&$5.163\pm0.694$&$1.21$\\
  &&H3&$-25.75\pm1.17$&$0.04$&$8.59\pm1.03$&$0.18$&$2.723\pm0.946$&$0.29$\\
  &\multirow{2}*{3}&H2&$-71.98\pm0.27$&$0.07$&$-71.09\pm0.28$&$0.39$&$1.175\pm0.098$&$1.79$\\
 &&H3&$-71.94\pm0.15$&$0.40$&$-71.10\pm0.15$&$0.67$&$2.938\pm0.162$&$0.38$\\
  \hline
  \multirow{6}*{TLoci}&\multirow{2}*{1}&H2&$52.96\pm0.17$&$0.35$&$2.59\pm0.12$&$0.92$&$2.883\pm0.175$&$0.67$\\
  &&H3&$53.10\pm0.23$&$0.87$&$2.62\pm0.16$&$0.50$&$2.201\pm0.147$&$1.35$\\
  &\multirow{2}*{2}&H2&$-25.82\pm0.21$&$0.10$&$8.55\pm0.15$&$1.00$&$5.602\pm0.394$&$1.01$\\
  &&H3&$-25.91\pm0.40$&$0.28$&$8.61\pm0.29$&$0.72$&$2.643\pm0.295$&$1.21$\\
  &\multirow{2}*{3}&H2&$-71.90\pm0.23$&$0.43$&$-71.00\pm0.22$&$0.91$&$0.993\pm0.082$&$0.09$\\
  &&H3&$-71.97\pm0.13$&$0.23$&$-71.17\pm0.13$&$0.23$&$3.016\pm0.166$&$0.10$\\
  \hline
  \multirow{6}*{Loci}&\multirow{2}*{1}&H2&$53.01\pm0.17$&$0.65$&$2.58\pm0.12$&$1.00$&$2.847\pm0.176$&$0.87$\\
  &&H3&$52.95\pm0.20$&$0.25$&$2.59\pm0.15$&$0.73$&$2.231\pm0.145$&$1.17$\\
  &\multirow{2}*{2}&H2&$-25.72\pm0.20$&$0.40$&$8.54\pm0.15$&$0.93$&$5.945\pm0.392$&$0.14$\\
  &&H3&$-25.76\pm0.36$&$0.11$&$8.63\pm0.27$&$0.85$&$2.952\pm0.294$&$0.16$\\
  &\multirow{2}*{3}&H2&$-71.91\pm0.22$&$0.41$&$-71.08\pm0.21$&$0.57$&$0.995\pm0.078$&$0.06$\\
  &&H3&$-71.95\pm0.13$&$0.38$&$-71.15\pm0.13$&$0.38$&$2.884\pm0.157$&$0.74$\\
  \hline
  \multirow{6}*{PCA5modes}&\multirow{2}*{1}&H2&$52.90\pm0.16$&$0.00$&$2.58\pm0.12$&$1.00$&$2.748\pm0.160$&$1.58$\\
  &&H3&$52.85\pm0.17$&$0.29$&$2.56\pm0.13$&$1.08$&$2.292\pm0.135$&$0.80$\\
  &\multirow{2}*{2}&H2&$-25.76\pm0.21$&$0.19$&$8.47\pm0.15$&$0.47$&$6.595\pm0.430$&$1.38$\\
  &&H3&$-25.94\pm0.45$&$0.31$&$8.56\pm0.33$&$0.48$&$3.432\pm0.361$&$1.20$\\
  &\multirow{2}*{3}&H2&$-72.00\pm0.24$&$0.00$&$-71.17\pm0.24$&$0.12$&$0.908\pm0.075$&$1.23$\\
  &&H3&$-71.92\pm0.13$&$0.62$&$-71.15\pm0.13$&$0.38$&$2.908\pm0.157$&$0.59$
 \end{tabular}
\caption{Measured astrometry and photometry in IRDIS images for each
  planet (same Id as in Table\,\ref{tab:fakepla}) using the model of
  planet image technique (section\,\ref{subsec:fit1}). Measurements are
  given with their $1\,\sigma$ uncertainties. For each measurement, we
  compare it to the true value using the $E$ criteria that is given in
  Eq.\,\ref{eq:E}.}
\label{tab:astrophoto}
\end{table*}
Each measurement $C$ is compared to the true value $C_R$ using
\begin{equation}
E=\frac{\left|C-C_R\right|}{Err},
\label{eq:E}
\end{equation}
with $Err$ the estimated $1\,\sigma$ uncertainty on $C$.

All the astrometric measurements are at less than $1\,\sigma$ from the
true values (Table\,\ref{tab:fakepla}), which corresponds to an
accuracy of $\sim0.2$\,pixel (i.e., $\sim2$\,mas); and all photometric
measurements are at less than $1.8\,\sigma$ from the true value $C_R$.
In the case of PCA images, the model of planet image technique can be
biased in the current version of SpeCal, especially when using more
than approximately ten modes. We are still investigating to understand why. To overcome this bias, we use the negative planet
technique (section\,\ref{subsec:fit2}) that can be time-consuming but which
provides more accurate measurements as showed in
Table\,\ref{tab:pca10modes}: the measurements are at less than
$1.5\,\sigma$ from the true values whereas they were at $6\,\sigma$
using the planet image technique.
\begin{table*}
\centering
\begin{tabular}{ccccc|cc|cc}
  Extraction&Id&band&$\Delta$RA(pixel)&E&$\Delta$DEC(pixel)&E&C$\times$1e6&E\\
   \hline
 \multirow{6}*{Model of planet image}&\multirow{2}*{1}&H2&$52.87\pm0.21$&$0.14$&$2.51\pm0.15$&$1.27$&$2.108\pm0.148$&$6.03$\\
  &&H3&$52.92\pm0.23$&$0.09$&$2.56\pm0.16$&$0.88$&$1.650\pm0.118$&$6.36$\\
  &\multirow{2}*{2}&H2&$-25.77\pm0.19$&$0.16$&$8.35\pm0.14$&$0.36$&$4.634\pm0.290$&$4.71$\\
  &&H3&$-26.00\pm0.33$&$0.61$&$8.34\pm0.21$&$0.29$&$2.182\pm0.204$&$4.01$\\
  &\multirow{2}*{3}&H2&$-72.07\pm0.23$&$0.30$&$-71.03\pm0.23$&$0.74$&$0.787\pm0.070$&$3.04$\\
 &&H3&$-71.93\pm0.15$&$0.47$&$-71.15\pm0.15$&$0.33$&$2.365\pm0.143$&$4.44$\\
  \hline
  \multirow{6}*{Negative planet}&\multirow{2}*{1}&H2&$52.90\pm0.10$&$0.00$&$2.58\pm0.07$&$1.71$&$2.781\pm0.146$&$1.50$\\
  &&H3&$52.84\pm0.12$&$0.50$&$2.56\pm0.08$&$1.75$&$2.298\pm0.126$&$0.81$\\
  &\multirow{2}*{2}&H2&$-25.74\pm0.10$&$0.60$&$8.47\pm0.08$&$0.88$&$6.338\pm0.321$&$1.05$\\
  &&H3&$-25.93\pm0.18$&$0.72$&$8.56\pm0.13$&$1.23$&$3.116\pm0.190$&$0.61$\\
  &\multirow{2}*{3}&H2&$-72.06\pm0.28$&$0.21$&$-71.28\pm0.19$&$0.42$&$0.915\pm0.086$&$0.99$\\
 &&H3&$-71.91\pm0.09$&$1.00$&$-71.13\pm0.09$&$0.78$&$2.919\pm0.162$&$0.50$
\end{tabular}
\caption{Extracted astrometry and photometry for each planet (same Id
  than in Table\,\ref{tab:fakepla}) using the model of planet image
  technique (section\,\ref{subsec:fit1}) and the negative planet technique
  (section\,\ref{subsec:fit2}) on the final image provided by PCA
  ten modes. For each measurement, we compare it to the true value using
  the $E$ criteria that is given in Eq.\,\ref{eq:E}.}
\label{tab:pca10modes}
\end{table*}

To conclude, SpeCal meets the requirements in terms of extracted
astrophysical signals because the measured astrometry and photometry
of point-like sources (e.g., exoplanets or brown dwarfs) are accurate
and unbiased. This is essential to correctly interpret the
observations (fit of the planet orbits or of their spectra). Moreover,
SpeCal enables the use of several algorithms to provide a global
dispersion of these measurements and a cross-check between measurements
to prevent any systematic errors.

\section{Reduction of IFS data}
\label{sec:exampleIfs}
We now use SpeCal to reduce one sequence recorded during
the SPHERE GTO on 2016 September 16 observing HD206893 in IFS YJH
mode. There are 80 images of $64$\,s exposure time and the
field of view rotates by 75.6 degrees. The seeing was about
0.7\,arcsec and the average wind speed was 8.4\,m.s$^{-1}$.

As for the IRDIS data, all algorithms are applied on the same datacube
provided by the first part of the SPHERE pipeline
\citep{pavlov08,mesa15,maireAL16b}. The datacube is a $290 \times 290
\times 80 \times 39$ array. The last dimension is the 
number of IFS spectral channels. We also added three fake planets to
the data (see Table \ref{tab:fakeplaIFS}) using the recorded stellar
PSFs. As for IRDIS, we chose the position and spectra to be
representative of three common cases. The spectra of planets 1 and 2
show strong variations between $0.9\,\mu$m and $1.7\,\mu$m. Planet 1
is closer to the star and its image is located in a region with bright
speckles. Planet 3 is located in a region with very bright speckles
and its spectrum in contrast is flatter than the others.
\begin{table}
\centering
\begin{tabular}{lcccc}
  Id&$\Delta$RA(pix)&$\Delta$DEC(pix)&Sepa(mas)&$C\times$1e6\\
  1&-36.70&21.20&316.2&4.3\\
  2&29.00&57.30&479.1&4.5\\
  3&2.40&-16.90&127.3&3.3
\end{tabular}
\caption{Fake planets injected in the IFS data: separation from the
  central star in pixels toward east ($\Delta$RA) and toward north
  ($\Delta$DEC), angular separation in mas, and averaged contrast over
  the $39$ spectral channels.}
\label{tab:fakeplaIFS}
\end{table}

\subsection{Calibration of the speckle pattern}
\begin{figure*}[!ht]
\centering
  \includegraphics[width=.6\textwidth]{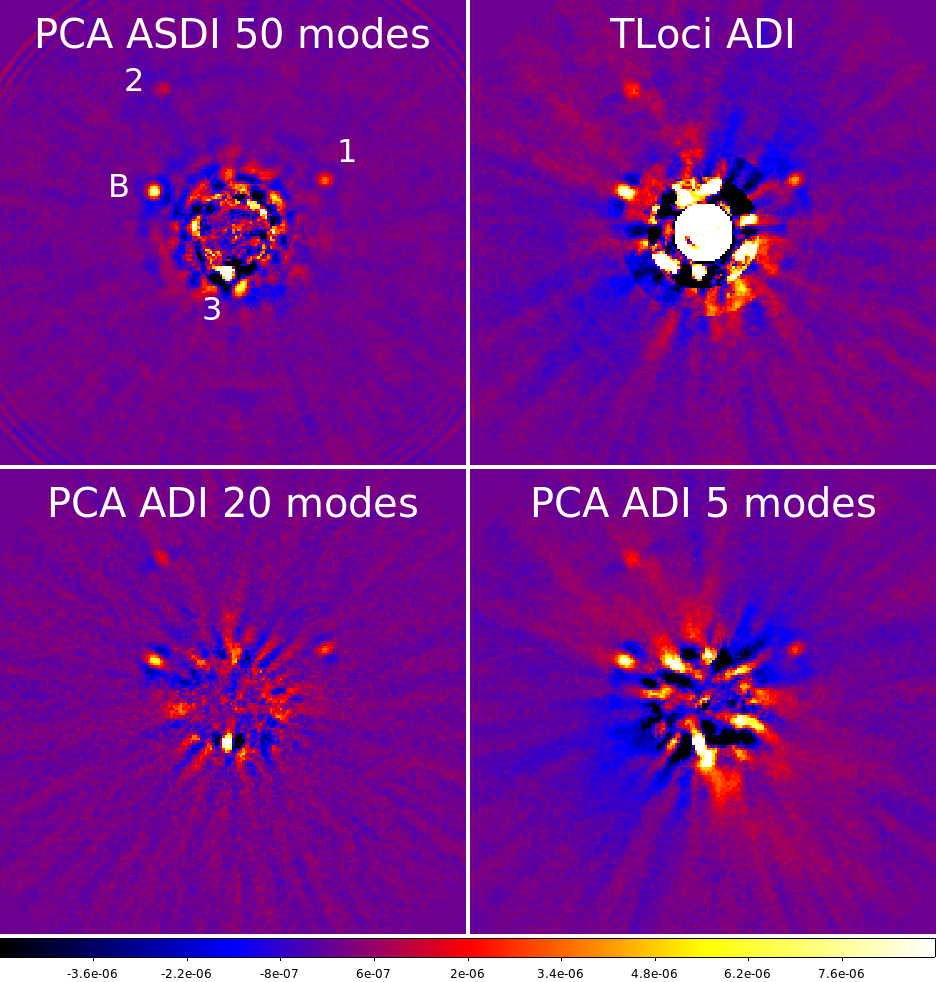}
  \caption{IFS example: Final images using PCA/ASDI (50 modes),
    PCA/ADI (20 and 5 modes), and TLoci. Images are corrected from the technique
    throughput and from the coronagraph transmission. The color scale.
    which is the same for all images, shows the contrast to the star
    ratio. The spatial scale is the same for all images.}
  \label{fig:imifs}
\end{figure*}

First, we apply ASDI PCA to detect point-like sources as it
efficiently minimizes the speckle pattern (Fig.\,\ref{fig:imifs}). We
detect four point-like sources: the three fake planets (1, 2, and 3) and
one real object HD\,206893\,B that we do not study in this paper
\citep[see][]{delormeP17,milli17}.
It is hard to accurately retrieve the planet photometry from ASDI images
when the planet spectrum is unknown, as demonstrated in
\citet{maireAL14} and \citet{rameau15}. Therefore, we also apply
several algorithms using ADI resulting in 39 final images
$I_{\mathrm{final}}(x,y,\lambda)$ for each algorithm. The averages
over the 39 channels are shown in~Fig.\,\ref{fig:imifs} for PCA/ADI
(5 and 20 modes) and TLoci/ADI. We clearly detect the four objects in
all images.

\subsection{Measurements of photometry}
For TLoci and PCA/ADI 5 mode images, we use the model of planet
image technique (section\,\ref{subsec:fit1}) to extract the photometry and
the astrometry of the three fake planets. For the PCA/ADI 20 mode
image, we use the negative planet technique (section\,\ref{subsec:fit2}) to
avoid the photometry underestimation that was noticed in
sectio\,\ref{subsec:astroIRDIS}. When considering the spectral channels
where the planet is detected, the astrometry measurements are accurate
with $1\,\sigma$ uncertainties of $0.6$, $0.3,$ and $0.5$ pixel for
planets $1$, $2$, and $3$ respectively. These values correspond to
$4.5$, $2.3,$ and $3.8$\,mas respectively.

\begin{figure}[!ht]
\centering
  \includegraphics[width=7.4cm]{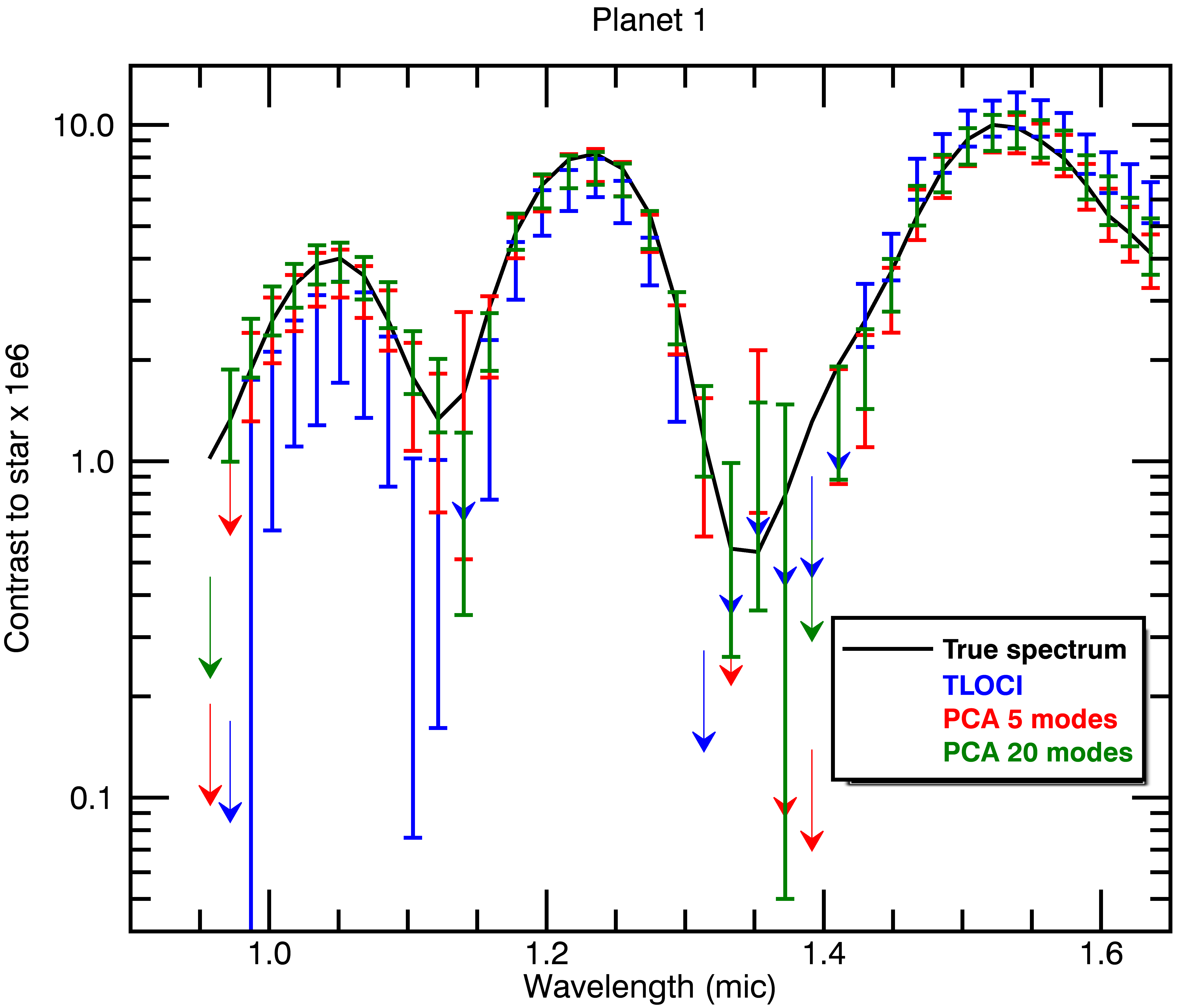}
  \caption{Spectra of planet-to-star contrast extracted from the
    PCA/ADI 5 modes (red), PCA/ADI 20 modes (green), and TLoci/ADI
    (blue) images compared to the true spectrum that was used for fake
    planet 1 (black full line). Error bars and upper limits are given
    at 1\,$\sigma$.}
  \label{fig:ifsspectrum1}
\end{figure}

\begin{figure}[!ht]
\centering
  \includegraphics[width=7.4cm]{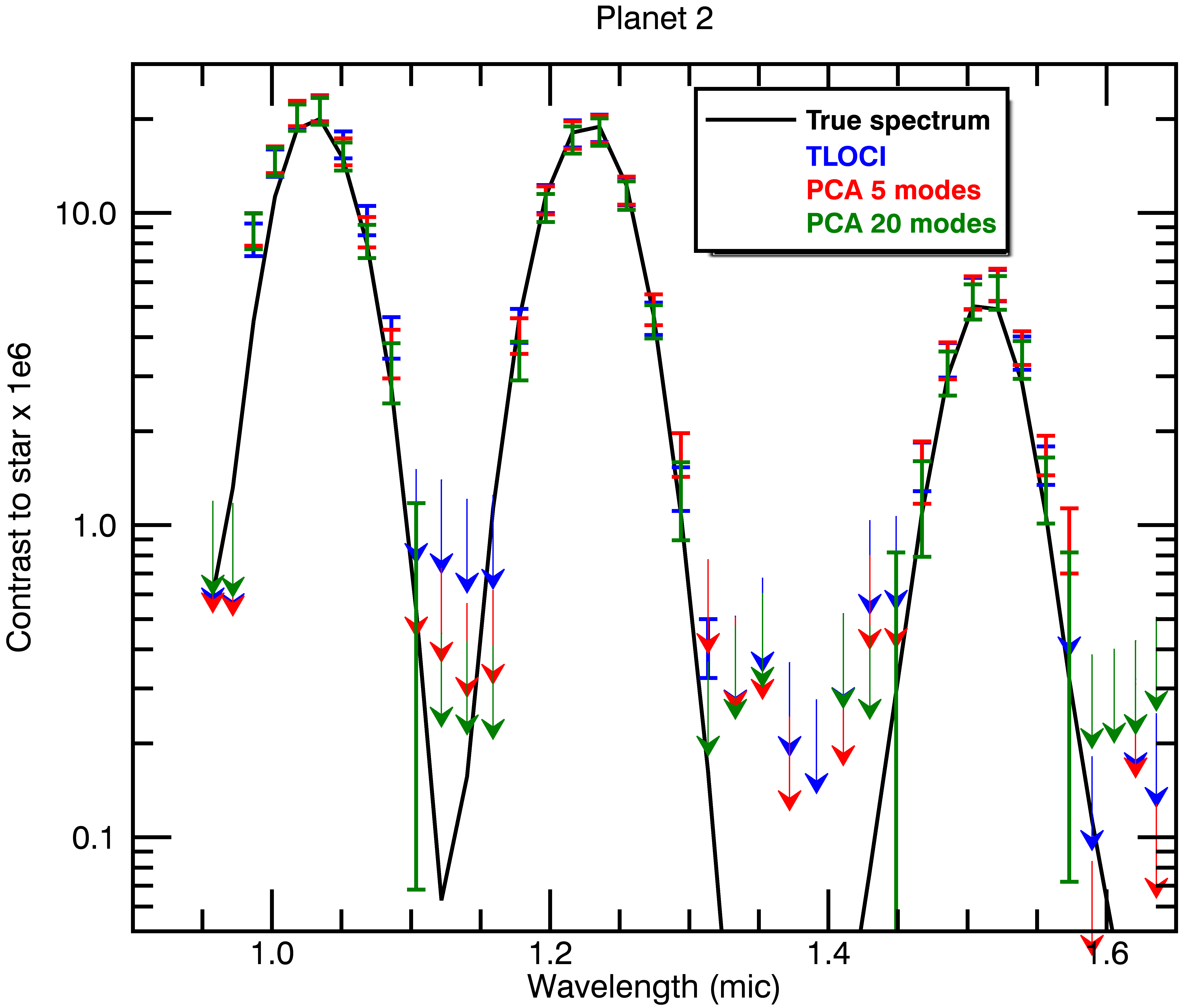}
  \caption{Same as Fig.\,\ref{fig:ifsspectrum1} for the fake planet
    2.}
  \label{fig:ifsspectrum2}
\end{figure}

\begin{figure}[!ht]
\centering
  \includegraphics[width=7.4cm]{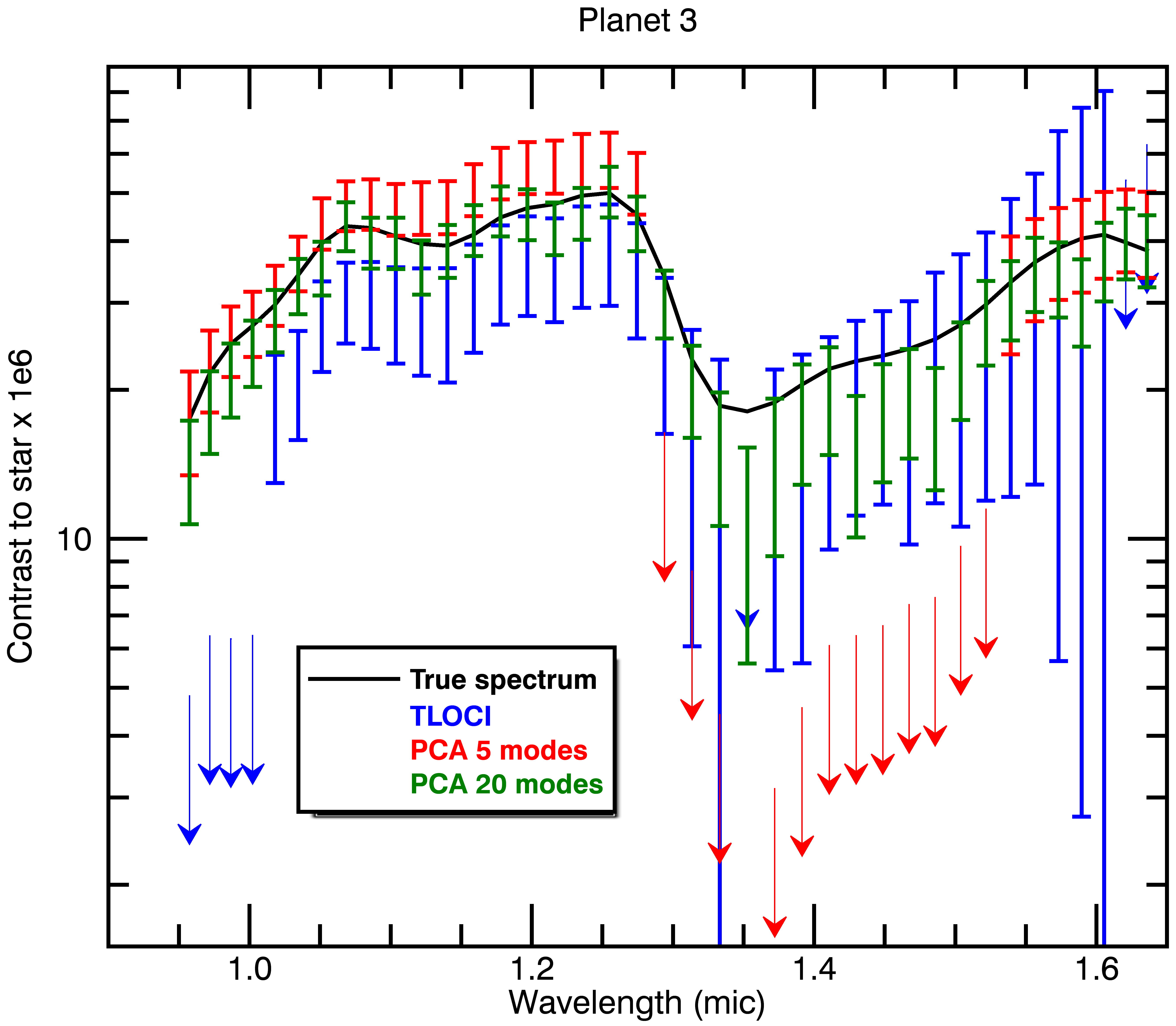}
  \caption{Same as Fig.\,\ref{fig:ifsspectrum1} for the fake planet
    3.}
  \label{fig:ifsspectrum3}
\end{figure}

We plot the extracted spectrophotometry in
Figs.\,\ref{fig:ifsspectrum1}, \ref{fig:ifsspectrum2}, and
\ref{fig:ifsspectrum3} for the TLoci/ADI (blue), the PCA/ADI 5 modes
(red), and the PCA/ADI 20 modes (green) cases, as well as the true
spectra that we used for the fake planets (full line).
The arrows give the $1\,\sigma$ upper limits when the planet is not
detected, and the error bars correspond to the estimated $1\,\sigma$
uncertainties.

All algorithms retrieve the spectra of the three fake planets with
similar uncertainties and no bias. In the case of planet 3, which is in
a region with bright speckles, there are spectral channels below
$1\,\mu$m for which TLoci give an upper limit only (no detection). For
the same planet, the PCA algorithm using 5-modes does not detect the object between
$1.3\,\mu$m  and $1.5\,\mu$m, whereas it does below
$1\,\mu$m. Such a situation often happens: several algorithms detect
the planet in different spectral channels. Thus, it is essential to
use several algorithms in parallel to optimize the detection and the
measured spectrum.

\section{Conclusion}
The Speckle Calibration tool (SpeCal) was developed by the SPHERE/VLT
consortium in the context of a large survey (SHINE), the main objective
of which is to search for and measure the astrometry and
spectrophotometry of exoplanets at large separations ($>5$\,au).
SpeCal provides high contrast images using a
variety of algorithms (cADI, PCA, Loci, TLoci) enabling the study of exoplanets, brown dwarfs, and circumstellar disks. SpeCal has been intensively tested on SPHERE guaranteed
time observations (GTO) and calibration data since 2013. It is
implemented in the SPHERE data center \citep{delormeP17b} to produce the final reduction for public data releases. The final
reductions will be available in the SPHERE target database
(TDB%\footnote{\href{http://cesam.lam.fr/spheretools/}%{http://cesam.lam.fr/spheretools/.}}
). Finally, SpeCal and
the DC are able to process all GTO data obtained with IRDIS/SPHERE
(dual-band imaging) and IFS/SPHERE (integral field spectrometer)
automatically.

SpeCal delivers major outputs for the survey and feeds the SPHERE
database with final images, contrast curves, signal to noise maps, astrometry, and
photometry of detected point-like sources in the field (exoplanets,
brown dwarfs, background sources, and all sub-stellar or stellar
candidates). This material has been used for the study of exoplanets
and circumstellar disks primarily based on SPHERE data
\citep{deboer16,ginsky16,lagrange16,maireAL16a,mesa16,perrot16,
  vigan16, zurlo16, benisty17,bonavita17,bonnefoy17,feldt17,maireAL17,
  mesa17, pohl17, samland17}.

In this paper, we investigated the astrometric and photometric
performance for point-like sources considering objects at a contrast
of $\sim3\times10^{-6}$ in the separation range of $3$ to $26$ \,{\sc
  fwhm}. Using the techniques of positive and negative fake planets, we
demonstrated the ability to achieve a measurement of the astrometry
with an accuracy of $\sim0.2$\,pixel (i.e., $\sim2$\,mas) for IRDIS and
$0.5$\,pixel (i.e., $\sim4$\,mas) for IFS. Similarly the photometric
accuracy reaches $\sim10\%$. 

We are planning to upgrade SpeCal with other algorithms like Andromeda
\citep{cantalloube15} or inverse approaches\,\citep{devaney17}, and
other observing modes (polarimetry, ZIMPOL). Finally, a specific tool
to model the ADI self-subtraction of circumstellar disks with simple
geometries will also be implemented.

\begin{acknowledgements}
 SPHERE is an instrument designed and built by a consortium consisting
 of IPAG (Grenoble, France), MPIA (Heidelberg, Germany), LAM
 (Marseille, France), LESIA (Paris, France), Laboratoire Lagrange
 (Nice, France), INAF -- Osservatorio di Padova (Italy), Observatoire
 astronomique de l'Universit\'e de Gen\`eve (Switzerland), ETH Zurich
 (Switzerland), NOVA (Netherlands), ONERA (France), and ASTRON
 (Netherlands), in collaboration with ESO. SPHERE was funded by ESO,
 with additional contributions from the CNRS (France), MPIA (Germany),
 INAF (Italy), FINES (Switzerland), and NOVA (Netherlands). SPHERE
 also received funding from the European Commission Sixth and Seventh
 Framework Programs as part of the Optical Infrared Coordination
 Network for Astronomy (OPTICON) under grant number
 RII3-Ct-2004-001566 for FP6 (2004--2008), grant number 226604 for FP7
 (2009--2012), and grant number 312430 for FP7 (2013--2016).
 
 This work has made use of the SPHERE data center, jointly operated by
 Osug/Ipag (Grenoble), Pytheas/Lam/Cesam (Marseille), OCA/Lagrange
 (Nice), and Observatoire de Paris/Lesia (Paris) and supported by a
 grant from Labex OSUG@2020 (Investissements d’avenir – ANR10
 LABX56).

 We acknowledge financial support from the French ANR GIPSE,
 ANR-14-CE33-0018.

Dino Mesa acknowledges support from the ESO-Government of Chile Joint
Committee program ''Direct imaging and characterization of exoplanets''.
\end{acknowledgements}
\bibliography{/home/raphael/biblio/bibtexbase/database.bib}   %>>>> bibliography data in report.bib
\bibliographystyle{aa}   %>>>> makes bibtex use spiebib.bst

\end{document}